\documentclass[10pt,journal,a4paper,twocolumn]{IEEEtran}
\usepackage{graphicx,color,amsmath,amsfonts,enumerate,amsthm,amssymb,mathtools,enumitem,thmtools,hyperref,subfigure,mathdots,enumitem,centernot,bm,soul,bbm,thm-restate}
\usepackage[capitalise, noabbrev]{cleveref}
\usepackage{multirow}
\hypersetup{colorlinks=true,linkcolor=blue,citecolor=blue,urlcolor=blue}
\usepackage{cite}

\def\C{ {\cal C} }
\def\D{ {\cal D} }
\def\E{ {\cal E} }
 
\def\G{ {\cal G} } 
 
\def\I{ {\cal I} }

\def\M{ {\cal M} }
\def\N{ {\cal N} }
  
\def\T{ {\cal T} }

\def\>{\rangle} 
\def\<{\langle}
\newcommand{\bra}[1]{\langle {#1} |}
\newcommand{\ket}[1]{| {#1} \rangle}
 
\newcommand{\ketbra}[2]{\ensuremath{\left|#1\right\rangle\!\!\left\langle#2\right|}}
\newcommand{\braket}[2]{\ensuremath{\!\!\left\langle#1|#2\right\rangle}\!\!}
\newcommand{\matrixel}[3]{\ensuremath{\left\langle #1 \vphantom{#2#3} \right| #2 \left| #3 \vphantom{#1#2} \right\rangle}}
\newcommand{\tr}[1]{\mathrm{Tr}\left( #1 \right)}

\newcommand{\iden}{\mathbbm{1}}

\renewcommand{\v}[1]{\ensuremath{\boldsymbol #1}}

\definecolor{ppblue}{RGB}{46,117,182}
\definecolor{ppgreen}{RGB}{46,182,117}
\definecolor{ppred}{RGB}{197, 90, 17}

\theoremstyle{plain}
\newtheorem{thm}{Theorem}

\newtheorem{cor}[thm]{Corollary}

\theoremstyle{definition}

\newtheorem{rmk}[thm]{Remark}

\begin{document}

\title{Encoding classical information \\into quantum resources}
\author{Kamil Korzekwa, Zbigniew Pucha{\l}a, Marco Tomamichel, Karol {\.Z}yczkowski
\thanks{Kamil Korzekwa is with International Centre for Theory of Quantum Technologies, University of Gda\'{n}sk, 80-308 Gda\'{n}sk, Poland, and jointly with Faculty of Physics, Astronomy and Applied Computer Science, Jagiellonian University, 30-348 Krak\'{o}w, Poland (email: korzekwa.kamil@gmail.com).}
\thanks{Zbigniew Pucha{\l}a is with Institute of Theoretical and Applied Informatics, Polish Academy of Sciences, 44-100 Gliwice, Poland, and jointly with Faculty of Physics, Astronomy and Applied Computer Science, Jagiellonian University, 30-348 Krak\'{o}w, Poland.}
\thanks{Marco Tomamichel is with the Centre for Quantum Technologies, National University of Singapore, Singapore 117543, Singapore, as well as with the Department of Electrical and Computer Engineering, National	University of Singapore, Singapore 117583, Singapore.	Part of this work was conducted while he was at the Centre for Quantum Software and Information, School of Software, University of Technology Sydney, Sydney NSW 2007, Australia.}
\thanks{Karol {\.Z}yczkowski is with Faculty of Physics, Astronomy and Applied Computer Science, Jagiellonian University, 30-348 Krak\'{o}w, Poland, and jointly with Center for Theoretical Physics, Polish Academy of Sciences, 02-668 Warszawa, Poland.}
\thanks{Copyright (c) 2017 IEEE. Personal use of this material is permitted.  However, permission to use this material for any other purposes must be obtained from the IEEE by sending a request to pubs-permissions@ieee.org.}}

\maketitle

\begin{abstract}
	We introduce and analyse the problem of encoding classical information into different resources of a quantum state. More precisely, we consider a general class of communication scenarios characterised by encoding operations that commute with a unique resource destroying map and leave free states invariant. Our motivating example is given by encoding information into coherences of a quantum system with respect to a fixed basis (with unitaries diagonal in that basis as encodings and the decoherence channel as a resource destroying map), but the generality of the framework allows us to explore applications ranging from super-dense coding to thermodynamics. For any state, we find that the number of messages that can be encoded into it using such operations in a one-shot scenario is upper bounded in terms of the information spectrum relative entropy between the given state and its version with erased resources. Furthermore, if the resource destroying map is the twirling channel over some unitary group, we find matching one-shot lower bounds as well. In the asymptotic setting where we encode into many copies of the resource state, our bounds yield an operational interpretation of resource monotones such as the relative entropy of coherence and its corresponding relative entropy variance.
\end{abstract}

\section{Introduction}
 \label{sec:intro}
 
Encoding information for storage or transmission is one of the most fundamental tasks in information theory~\cite{shannon48}. A typical communication scenario in which a $d$-dimensional quantum system is employed to transfer classical information between a sender $S$ and a receiver $R$ is composed of three stages. First, $S$ encodes a message $m\in\{1,\dots,M\}$ by preparing a quantum system in a state~$\rho_m$; then, she sends it to $R$ via a noisy quantum channel~$\N$; finally, $R$ decodes the original message by performing a measurement on $\N(\rho_m)$. Crucially, it is assumed that the encoding and decoding steps are unconstrained, so that the problem of optimal information transfer (finding the maximal $M$ for a given average decoding error~$\epsilon$) is fully specified by the noisy channel $\N$. Holevo, Schumacher and Westmoreland~\cite{holevo98,schumacher97} analyzed this problem in an asymptotic setting where $\N$ is memoryless and can be used multiple times. They established the regularized Holevo information~\cite{holevo73b} of $\N$ as the maximal transmission rate (in bits per channel use) that still allows for asymptotically vanishing error.\footnote{Here one considers the setting where a joint decoding measurement is allowed, but no entanglement between channel inputs is present, avoiding the issue of super-additivity~\cite{hastings09}.} More recently, various refinements of the trade-off between the decoding error and the transmission rate have been established~\cite{ogawa99,winter99,hayashi03,mosonyi2009generalized,wang10,tomamicheltan14,mosonyi2017strong,chubb17,cheng17,dalai13,cheng19} when the number of channel uses is finite.  

Here, we analyze an alternative communication scenario, where we assume that $S$ and $R$ are connected via a noiseless channel, $\N=\I$, but the encoding ability of $S$ is constrained. In particular, we assume that $S$ does not have the ability to prepare arbitrary quantum states. Instead, she is given a quantum system in a state $\rho$ that acts as an information carrier, and can encode a message $m$ into it by applying an encoding $\E_m$ from a constrained set of quantum channels. We focus on a family of constraints, each of which is defined via a fixed idempotent channel $\D$ and the following conditions that all allowed encoding maps $\E_m$ have to satisfy,
\begin{subequations}
\begin{align}
	\label{eq:constr_1}
	\E_m\circ \D &=  \D,\\
	\label{eq:constr_2}
	\D \circ \E_m &= \D.
\end{align}
\end{subequations}
Channels $\E_m$ satisfying Eqs.~\eqref{eq:constr_1}-\eqref{eq:constr_2} will be called \emph{encodings into resources destroyed by $\D$}. 

To understand the meaning of such constraints, note that one can interpret $\D$ as a resource destroying map~\cite{liu2017resource} that erases information encoded in degrees of freedom that $\E_m$ affects. More precisely, application of $\D$ to any state $\rho$ renders it useless from the perspective of $S$, as $\D(\rho)$ is invariant under $\E_m$ through Eq.~\eqref{eq:constr_1}. Similarly, application of $\D$ to an encoded state $\E_m(\rho)$ renders it useless from the perspective of $R$, as every $\E_m(\rho)$ gets sent to a fixed state $\D(\rho)$ according to Eq.~\eqref{eq:constr_2}. In other words, the above conditions restrict $S$ to encodings satisfying resource non-generating and non-activating conditions~\cite{liu2017resource}, which additionally cannot modify free (non-resource) states. Therefore, by focusing on encodings into resources, one investigates the capability of particular degrees of freedom, corresponding to resources destroyed by $\D$, to carry classical information (note here that resource-destroying channels exist only for a subset of convex resource theories~\cite{liu2017resource,gour2017quantum}). Alternatively, Eq.~\eqref{eq:constr_1} tells us that no communication is possible if Alice is given a state in the image of $\D$, while Eq.~\eqref{eq:constr_2} tells us that no communication is possible if Bob can only use measurements with measurement operators in the image of $\D^*$.

Although specifying the encoding constraints via a channel $\D$ may initially look a little abstract, the plethora of scenarios that can be captured this way justifies the definition. One of the particularly useful choices of $\D$ is the $G$-twirling channel $\G$ over some subgroup $G$ of unitary channels, i.e.,
\begin{equation}
	\label{eq:twirling}
	\G(\rho):=\frac{1}{|G|}\sum_{U_g\in G} U_g\rho U_g^\dagger,
\end{equation}
where for continuous groups the above sum can be replaced by an integral $\int d\mu(U_g) $ with respect to the Haar measure. An example of encodings that satisfy conditions from Eq.~\eqref{eq:constr_1}-\eqref{eq:constr_2} for a $G$-twirling channel is given by unitary channels $U_g(\cdot)U_g^\dagger$ belonging to $G$. This is due to the fact that for general groups $G$ one has \mbox{$\forall h\in G:~hG=Gh=G$}, and so
\begin{equation}
	\label{eq:rearrangement}
	\forall U_g:\quad \G(U_g(\cdot)U_g^\dagger)=U_g\G(\cdot)U_g^\dagger=\G(\cdot).
\end{equation}
Three important comments concerning encodings into resources destroyed by the $G$-twirling channel~$\G$ are now in place. First, all such encodings must be unital, which comes directly from Eq.~\eqref{eq:constr_1} and the fact that $\G$ is unital. Second, smaller subgroups yield stronger restrictions on the set of applicable encodings, i.e., if $G_1\subset G_2$, then \mbox{$\E_m\circ\G_1=\G_1\circ\E_m=\G_1$} implies \mbox{$\E_m\circ\G_2=\G_2\circ\E_m=\G_2$}. And finally, the considered encodings should not be confused with group covariant encodings (see Appendix~\ref{app:covariant} for a detailed discussion of that issue).

When $G$ is a full unitary group, $\G$ becomes the completely depolarizing channel, i.e., $\G(\rho)=\iden/d$ for all $\rho$. Then, Eq.~\eqref{eq:constr_2} is satisfied automatically and Eq.~\eqref{eq:constr_1} constrains the encoding to unital channels. Physically, this case corresponds to the sender $S$ being unable to decrease the entropy of the information carrier, and so the information is encoded into the resource of purity~\cite{horodecki2013quantumness}. Moreover, if $G$ is the unitary group on a subsystem $S_1$ of a multipartite system $S_{1 \dots n}$, the encodings are restricted to unital channels acting locally on $S_1$. This way one can study encoding information not only into a resource of local purity, but also in entanglement, allowing one to assess the capacity of the system for super-dense coding~\cite{bennett1992communication}. 

When $G$ is a subgroup of all unitaries diagonal in a given basis $\{\ket{k}\}$ (so it is a subgroup of commuting unitaries), $\G$ becomes the completely dephasing map $\Delta$ with respect to this basis, i.e., \mbox{$\Delta(\rho)=\sum_k \bra{k}{\rho}\ket{k}\ketbra{k}{k}$}. Equations~\eqref{eq:constr_1}-\eqref{eq:constr_2} then constrain encoding maps $\E_m$ to be Schur-product channels~\cite{kye1995positive,li1997special}, i.e., \mbox{$\E_m(\rho)=\rho\circ C_m$} with $C_m$ being arbitrary correlation matrices ($C_m\geq 0$ and all diagonal entries being 1), and $\circ$ denoting Schur (entry-wise) product (not to be confused with composition of quantum channels as in Eqs.~\eqref{eq:constr_1}-\eqref{eq:constr_2}). Since such channels do not modify populations (diagonal elements of $\rho$ in the given basis), but only affect coherences (off-diagonal elements), investigating communication scenarios with this constraint corresponds to asking how much classical information can be encoded into the resource of quantum coherence~\cite{aberg2006superposition,baumgratz2014quantifying}. Finally, for a general group $G$, the channel $\G$ is a projector onto symmetric states (i.e., states invariant under the action of all group elements), and so $\G$-constrained encodings correspond to sending information encoded in the resource of asymmetry~\cite{marvianthesis}, i.e., into the degrees of freedom that are not invariant under the group action.

Another important choice of $\D$, besides that of the $G$-twirling channel, is given by the completely thermalising map, i.e., $\T(\rho)=\gamma$ for all $\rho$, with $\gamma$ denoting the thermal equilibrium Gibbs state. In this case, the considered constraints limit the sender $S$ to only encode information with Gibbs-preserving channels, i.e., satisfying $\E_m(\gamma)=\gamma$. Physically, this constrains $S$ to obey the second law of thermodynamics (in the sense that the encoding channels cannot bring the information carrier farther out of equilibrium), and the information is encoded in the resource of thermodynamic non-equilibrium~\cite{brandao2013resource,brandao2015second}. 

In this paper, we first derive a single-shot upper bound for the number of classical messages that can be encoded in quantum resources and then decoded up to average probability of error $\epsilon$. We then also derive a single-shot lower bound for the particular case of a resource destroying map $\D$ given by the $G$-twirling channel. Next, we show how, in the limit of a large number of information carriers, the two bounds coincide, yielding the optimal encoding rate up to the second order asymptotic expansion. These rates are given by the relative entropy and relative entropy variance between the state of the information carrier $\rho$ and the same state with erased resource content $\G(\rho)$. We thus provide an operational meaning to a number of resource monotones used in various resource theories. In what follows, we first formally state and prove our results in Sec.~\ref{sec:result}. Then, in Sec.~\ref{sec:discussion}, we discuss their relevance and applications in a variety of quantum communication scenarios. Finally, we present an outlook for future research in Sec.~\ref{sec:outlook}.

\section{Main results}
\label{sec:result}

\subsection{Formal statement of the problem}

Our main goal is to encode a message $m$, chosen uniformly at random from the set $\M=\{1,\dots,M\}$, using a $d$-dimensional quantum state $\rho$ and a constrained set of quantum channels, described by Eqs.~\eqref{eq:constr_1}-\eqref{eq:constr_2}, so that this message can be faithfully recovered later up to average probability of error $\epsilon$. We are thus looking for an encoder in terms of a set of quantum channels $\{\E_m\}_{m\in\M}$ that are encodings into resources destroyed by some fixed channel $\D$; and a decoder specified by a quantum measurement described by the POVM elements $\{D_m\}_{m\in\M}$, such that
\begin{equation}
	\label{eq:distingushability_condition}
	\frac{1}{M}\sum_{m=1}^M\tr{\E_m(\rho)D_m}\geq 1-\epsilon,
\end{equation}
i.e., the average probability of incorrectly decoding the message is smaller then $\epsilon$. We are interested in the maximal allowed value of $M$ for given $\rho$ and $\epsilon$. Of special importance is the case when we deal with $N$ independent and identically distributed copies of a state $\rho$, i.e., when we encode into $\rho^{\otimes N}$ using encodings into resources destroyed by $\D^{\otimes N}$. Then, the aim is to find the optimal rate $R$ (in bits per resource use) for encoding information into resources of quantum states,
\begin{align}
\label{eq:rate}
R(\rho,N,\epsilon):=\sup \left\{\frac{\log M}{N} \ \big| \ \textrm{Eq.~\eqref{eq:distingushability_condition}~holds}\right\},
\end{align}
and, in particular, to understand its asymptotic behavior as $N\rightarrow\infty$. Here and throughout, $\log$ refers to the binary logarithm.

In order to present our results, we first need to introduce several entropic quantities. The relative entropy $D$ between density matrices $\rho$ and $\sigma$ is defined by~\cite{umegaki62}
\begin{equation}
D(\rho \| \sigma) :=\operatorname{tr}\left(\rho(\log \rho-\log \sigma)\right),
\end{equation}
while the relative entropy variance $V$ is given by~\cite{tomamichel12,li2014second}
\begin{equation}
V(\rho \| \sigma) :=\operatorname{tr}\left(\rho(\log \rho-\log \sigma)^{2}\right)-D(\rho \| \sigma)^2.
\end{equation}
These quantities will be relevant for the statement of our asymptotic results.

Our single-shot bounds are given in terms of two tightly related quantities: the information spectrum relative entropy and the hypothesis testing relative entropy. The information spectrum relative entropy, $D_s^\delta(\cdot\|\cdot)$, is given as~\cite{hayashi03}
\begin{equation}
\label{eq:spectrum_rel_ent}
D_s^\delta(\rho\|\sigma):=\sup \big\{K\ \big| \ \tr{\rho \Pi_{\rho\leq 2^K \sigma}}\leq \delta\big\},
\end{equation}
where $\Pi_{\rho\leq 2^K \sigma}$ is the orthogonal projection onto the subspace generated by eigenspaces of $2^K \sigma-\rho$ with non-negative eigenvalues.
Moreover, the hypothesis testing relative entropy, $D_H^{\epsilon}(\cdot\|\cdot)$, is defined as~\cite{wang10,buscemi2010quantum,brandao2011one} 
\begin{align}
D_H^{\epsilon}(\rho\|\sigma) := - \log \inf \big\{ \tr{A\sigma}\ \big|\ &0 \leq A \leq \iden,\nonumber\\
&\tr{A\rho} \geq 1-\epsilon \big\} \,.
\label{eq:hypothesis_rel_ent}
\end{align}
These two quantities are equivalent up to asymptotically vanishing terms~\cite[Lemma 12]{tomamichel12}, namely
\begin{align}
D_s^{\epsilon} ( \rho \| \sigma ) \leq  D_H^{\epsilon}( \rho \| \sigma ) \leq D_s^{\epsilon+\delta}( \rho \| \sigma ) + \log \frac1\delta \,,
\label{eq:lemma12}
\end{align}
where $\epsilon\in[0,1)$ and $\delta\in(0,1-\epsilon)$. The hypothesis testing relative entropy is operational and measures the ultimate trade-off between errors of the first and second kind in binary hypothesis testing. It satisfies a data-processing inequality, i.e., for a positive trace-preserving map $\mathcal{E}$, we have $D_H^{\epsilon}( \rho \| \sigma ) \geq D_H^{\epsilon}( \mathcal{E}(\rho) \| \mathcal{E}(\sigma) )$. The information spectrum relative entropy, on the other hand, is a non-commutative generalisation of a tail bound of the cumulative distribution of the log-likelihood ratio, and is operationally relevant only in the asymptotic limit. We refer to Ref.~\cite{tomamichel12} for a further discussion of these quantities.\footnote{See also Ref.~\cite{leditzky16} for an alternative definition of the information spectrum relative entropy that satisfies data-processing.}

\subsection{One-shot bounds}

We have the following upper bound for the number of messages $M$ that can be encoded in resources erased by an arbitrary resource destroying map $\D$.
\begin{restatable}[Single-shot upper bound]{lem}{lemmaone}
	\label{lemma:upper}
	The number of messages $M(\rho,\epsilon)$ that can be encoded into resources of a quantum state $\rho$ destroyed by $\D$, for an average decoding error at most $\epsilon$, is upper bounded as follows:
	\begin{equation}
		\label{eq:upper}
		\log M(\rho,\epsilon)\leq D_H^{\epsilon}( \rho \| \D(\rho)) \,.
	\end{equation}
\end{restatable}

The proof relies on a by now standard technique for relating information-theoretic problems with hypothesis testing, which in the quantum setting can be traced back at least to Ref~\cite{hayashi03}, while our notation follows Ref.~\cite{wang10}.

\begin{proof}
First, we consider the final state after the encoding and correlated with the chosen message,
\begin{align}
	\tau_{MQ} =  \frac{1}{M}\sum_{m \in \M} \ketbra{m}{m} \otimes \E_m(\rho).
\end{align}
We now assume that there exists a suitable decoder $\{ D_m \}_{m\in\M}$ for a faithful recovery up to average failure probability $\epsilon$ (i.e., Eq.~\eqref{eq:distingushability_condition} is fulfilled), and will show that this leads to an upper bound on $M$ given by Lemma~\ref{lemma:upper}. To prove this, let us take a closer look at the hypothesis testing relative entropy $D_H^{\epsilon}$ between $\tau$ and
\begin{equation}
	\zeta := \frac{1}{M}\sum_{m \in \M}  \ketbra{m}{m} \otimes \D(\rho) \,.
\end{equation}
By recalling the definition of $D_H^{\epsilon}$, Eq.~\eqref{eq:hypothesis_rel_ent}, we see that 
\begin{equation}
	D_H^{\epsilon}(\tau_{MQ}\|\zeta)\geq -\log \tr{A\zeta}
\end{equation}
for
\begin{equation}
	A = \sum_{m \in \M} \ketbra{m}{m} \otimes D_m.
\end{equation}
This is because the above (potentially suboptimal) choice of $A$ clearly satisfies \mbox{$0\leq A\leq \iden$} and also
\begin{equation}
	\tr{A\tau_{MQ}}=	\frac{1}{M} \sum_{m \in \M} \tr{\E_m(\rho)D_m}\geq 1-\epsilon,
\end{equation} 
with the final inequality holding, because we assumed that Eq.~\eqref{eq:distingushability_condition} is fulfilled. At the same time we have
\begin{equation}
	\tr{A\zeta}=\frac{1}{M} \sum_{m \in \M} \tr{\D(\rho)D_m}=\frac{1}{M},
\end{equation} 
so that 
\begin{equation}
	\label{eq:hypothesis_ineq_1}
	\log M\leq D_H^{\epsilon}(\tau_{MQ}\|\zeta).
\end{equation}

Next, we employ the data-processing inequality twice. First, for the channel
\begin{equation}
	\tilde{\E}:=\sum_{m \in \M} \ketbra{m}{m} \otimes \E_m
\end{equation}
that leaves $\zeta$ unchanged, and then for tensoring with the maximally mixed state of the message register. This yields the following sequence of inequalities:
\begin{align}
	D_H^{\epsilon}(\tau_{MQ}\|\zeta)&=D_H^{\epsilon}\left(\tilde{\E}\left(\frac{1}{M}\sum_{m \in \M} \ketbra{m}{m} \otimes \rho\right)\bigg\|\tilde{\E}(\zeta)\right)\nonumber\\
	&\leq D_H^{\epsilon}\left(\frac{1}{M}\sum_{m \in \M} \ketbra{m}{m} \otimes \rho\bigg\|\zeta\right)\nonumber\\
	&\leq D_H^{\epsilon}\big( \rho \big\| \D(\rho) \big).
\end{align}
Combining this with Eq.~\eqref{eq:hypothesis_ineq_1}, we finally arrive at
\begin{equation}
	\label{eq:hypothesis_ineq_2}
	\log M\leq D_H^{\epsilon}\big( \rho \big\| \D(\rho) \big).
\end{equation}

\end{proof}

\begin{rmk}
	Lemma~\ref{lemma:upper} holds more generally for all encodings satisfying only Eq.~\eqref{eq:constr_1} and not necessarily Eq.~\eqref{eq:constr_2}, since in the proof the latter assumption was not used. Moreover, the requirement of $\D$ being idempotent is not necessary.
\end{rmk}

For a resource destroying map given by the $G$-twirling channel $\G$, the following theorem bounds $M$ in the single-shot setting from both sides.\footnote{Lemma~\ref{lemma:upper} gives an alternative, slightly tighter, upper bound; however, we opted here for a more symmetrical exposition.}

\begin{restatable}[Single-shot encoding]{thm}{theoremtwo}
	\label{thm:single-shot}
	The maximum number of messages $M(\rho,\epsilon)$ that can be encoded into resources of a quantum state $\rho$ destroyed by the $G$-twirling channel $\G$ over a unitary subgroup $G$, for an average decoding error at most $\epsilon$, is bounded by
	\begin{subequations}
	\begin{align}
		\label{eq:bounds1}
		\log M(\rho,\epsilon)&\leq D_s^{\epsilon+\delta}( \rho \| \G(\rho)) + \log \frac1\delta\, ,\\
		\label{eq:bounds2}
		\log M(\rho,\epsilon)&\geq D_s^{\epsilon-\delta}(  \rho\| \G(\rho)) - \log \frac{2}{\delta}\, ,
	\end{align}
	\end{subequations}
	for all $\delta\in(0, \min\{\epsilon, 1-\epsilon\})$. Moreover, $M(\rho,\epsilon)$ can be attained using unitary encodings belonging to $G$.
\end{restatable}

\begin{rmk}
	\label{rmk:cq-channel}
	Theorem~\ref{thm:single-shot} can be seen as a generalization of Lemmas~25~and~20 of Ref.~\cite{hayashi2020finite}. More precisely, if we consider a classical-quantum channel
	\begin{equation}
		\label{eq:cq_channel}
		g\in G \mapsto \rho_g:= U_g \rho U_g^\dagger,
	\end{equation}
	then Eq.~\eqref{eq:bounds2} can be understood as its one-shot achievability bound, the classical special case of which was investigated in Refs.~\cite{verdu1994general,hayashi2020finite}. Similarly, Eq.~\eqref{eq:bounds1} can be understood as a one-shot converse bound. It is, however, more general even in the classical case, as it considers any encodings into resources destroyed by $\G$, of which the encoding from Eq.~\eqref{eq:cq_channel} is a special case. Finally, we note that our achievability proof follows steps analogous to the ones presented in Ref.~\cite{beigi2014quantum}.	
\end{rmk}

\begin{proof}
	
The upper bound can be proven by employing Lemma~\ref{lemma:upper}. We simply note that according to Eq.~\eqref{eq:lemma12}, we have
\begin{align}
D_H^{\epsilon}\big( \rho \big\| \D(\rho) \big) \leq D_s^{\epsilon+\delta}\big( \rho \big\| \D(\rho) \big) + \log \frac1\delta,
\end{align} 
for $\delta \in (0, 1-\epsilon)$, which yields the upper bound in Theorem~\ref{thm:single-shot}.

For the lower bound (achievability), we will consider a family of encoder-decoder pairs, each defined by its codebook~$\C$. We will then show that by choosing the encoder-decoder pair from this family uniformly at random, and encoding the number of messages given by the claimed lower bound, the expected probability of error is below the required threshold. This, in turn, implies that within the introduced family there must exist at least one encoder-decoder pair that allows for encoding that number of messages and decoding with error at most $\epsilon$. 

A given codebook $\C$ is defined by a mapping from the set of messages $\M$ to the set of unitaries $\{U_{g_1},\dots,U_{g_M}\}$ from the group $G$, so that the message $m$ is encoded into the following quantum state:
\begin{equation}
	\rho_{m} = U_{g_m}\rho U_{g_m}^\dagger.
\end{equation}
The decoder, on the other hand, is given by a pretty good measurement consisting of $M$ POVM elements $D_m$ defined by
\begin{equation}
	D_m:= \frac{1}{M}\omega^{-1/2}\rho_{m}  \omega^{-1/2},
\end{equation}	
with
\begin{equation}
	\omega=\frac{1}{M}\sum_{m=1}^M \rho_{m}.
\end{equation}

The probability $p_s(\C)$ of successfully decoding the message encoded with the use of a codebook $\C$ is then given by
\begin{align}
	p_s(\C)&=\frac{1}{M}\sum_{m=1}^M \tr{\rho_{m} D_m}\nonumber\\
	&=\frac{1}{M^2}\tr{\sum_{m=1}^M\left(\omega^{-1/4}\rho_m\omega^{-1/4}\right)^2}\nonumber\\
	&=\frac{1}{M^2}\sum_{m =1}^M Q_2(\rho_m\|\omega),
\end{align} 
where
\begin{equation}
	Q_2(\rho\|\sigma)=2^{D_2(\rho\|\sigma)}
\end{equation}
with the collision relative entropy $D_2$ defined by~\cite{renner2008security}
\begin{align}
	\label{eq:collision_rel_ent}
	D_2(\rho\|\sigma):=\log \tr{\left(\sigma^{-1/4}\rho \sigma^{-1/4}\right)^2}.
\end{align}

We now consider a random choice of the codebook $\C$: each message $m$ is independently encoded into a group element $U_{g_m}$ uniformly at random, i.e., all $U_{g_m}$ are independent and identically distributed according to the Haar uniform measure $d\mu(U_{g_m})$. The success probability $P_s$ averaged over all choices of codebooks is then given by
\begin{align}
	P_s&:=\int d\mu(U_{g_1})\dots d\mu(U_{g_M}) p_s(\C)\nonumber\\
	&=\frac{1}{M^2}\sum_{m =1}^M \int d\mu(U_{g_1})\dots d\mu(U_{g_M})   Q_2(\rho_m\|\omega)\nonumber\\
	&\geq \frac{1}{M^2} \sum_{m=1}^M\int  d\mu(U_{g_m}) Q_2\left(\rho_m\middle\|\frac{1}{M}\rho_m + \frac{M-1}{M} \G(\rho)\right)\nonumber\\
	&= \frac{1}{M^2} \sum_{m=1}^M\int  d\mu(U_{g_m}) Q_2\left(\rho\middle\|\frac{1}{M}\rho + \frac{M-1}{M} \G(\rho)\right)\nonumber\\
	&= \frac{1}{M} Q_2\left(\rho\middle\|\frac{1}{M}\rho+\frac{M-1}{M}\G(\rho)\right).	
\end{align} 
Here, we first used the convexity of $Q_2$ in the second argument~\cite{muller2013quantum} together with the definition of $G$-twirling $\G$, and then we employed the unitary invariance of $Q_2$.

Now, we use Theorem~3 of Ref.~\cite{beigi2014quantum}, that allows us to lower bound the above expression by replacing $D_2$ with the information spectrum relative entropy $D_s^{\delta'}$, defined in Eq.~\eqref{eq:spectrum_rel_ent}. Namely, for $\delta' = \epsilon - \delta$ such that $\delta' \in (0, \epsilon)$, it holds that
\begin{align}
	&2^{D_2( \rho\| \frac{1}{M}\rho+\frac{M-1}{M}\G(\rho))}\nonumber\\
	&\qquad\geq (1-{\delta'})\left(\frac{1}{M}+\frac{M-1}{M}2^{-D_s^{{\delta'}}( \rho\| \G(\rho))}\right)^{\!-1}\nonumber\\
	&\qquad\geq M(1-{\delta'})\left(1-(M-1)2^{ -D_s^{{\delta'}}( \rho\| \G(\rho))}\right)\nonumber\\
	&\qquad\geq M(1-{\delta'})(1-2^{ -D_s^{{\delta'}}( \rho\| \G(\rho))}M).
\end{align} 
We can use the above to bound $P_s$ as follows,
\begin{align}
P_s\geq (1-{{\delta'}}) (1-2^{-D_s^{{\delta'}}( \rho\| \G(\rho))}M).
\end{align}
Therefore, if we choose
\begin{align}
M = \left \lfloor \frac{\epsilon-{{\delta'}}}{1-{{\delta'}}} 2^{D_s^{{\delta'}}( \rho\| \G(\rho))} \right \rfloor ,
\end{align}
we find
\begin{align}
P_s \geq (1-{{\delta'}}) (1-2^{-D_s^{{\delta'}}( \rho\| \G(\rho))}M) \geq 1-\epsilon,
\end{align}
as required. Next, note that the bound in Eq.~\eqref{eq:bounds2} is trivially satisfied when 
\begin{equation}
	2^{-D_s^{\delta'}( \rho\| \G(\rho))}\geq  \frac{\delta}{2},
\end{equation}
and it remains to consider the case \mbox{$2^{-D_s^{\delta'}( \rho\| \G(\rho))}\leq \frac{\delta}2$}. Then,
\begin{align}
\log M(\rho,\epsilon) &\geq \log M \nonumber\\
&\geq \log \left( \frac{\delta}{1-\delta'} 2^{D_s^{\delta'}( \rho\| \G(\rho))} - 1 \right) \nonumber\\
&= D_s^{\delta'}(  \rho\| \G(\rho)) - \log \frac{1}{\delta} \nonumber\\
&\qquad +  \log \left(  \frac{1}{1-\delta'} - \frac{ 2^{-D_s^{\epsilon-\delta}( \rho\| \G(\rho))} }{\delta}  \right) \nonumber\\
&\geq D_s^{\delta'}(  \rho\| \G(\rho)) - \log \frac{2}{\delta}\,,
\end{align}
which implies the desired inequality.
\end{proof}

\begin{rmk}
	Using Lemma~\ref{lemma:upper} and Eq.~\eqref{eq:lemma12}, one can rewrite the bounds from Theorem~\ref{thm:single-shot} in terms of the hypothesis testing relative entropy:	
		\begin{subequations}
		\begin{align}
			\label{eq:bounds1hyp}
			\log M(\rho,\epsilon)&\leq D_H^{\epsilon}( \rho \| \G(\rho)),\\
			\label{eq:bounds2hyp}
			\log M(\rho,\epsilon)&\geq D_H^{\epsilon-2\delta}(  \rho\| \G(\rho)) - \log \frac{2}{\delta^2}\,,
		\end{align}
	\end{subequations}
 with $\epsilon\in[0,1)$ and $\delta\in(0,\epsilon/2)$.
\end{rmk}

\subsection{Asymptotic expansion}

From now on we will use the asymptotic notation, with $\simeq$, $\lesssim$ and $\gtrsim$ denoting equalities and inequalities up to terms of the order $O(\log N/N)$. The asymptotic encoding rate into quantum resources destroyed by $\G$ is captured by the following theorem:
\begin{restatable}[Asymptotic encoding]{thm}{theoremthree}
	\label{thm:asymptotic}
	The optimal rate $R(\rho,N,\epsilon)$ for encoding information into resources of a quantum state $\rho^{\otimes N}$ destroyed by the $G$-twirling channel $\G^{\otimes N}$ over a unitary subgroup $G^{\times N}$, for an average decoding error at most~$\epsilon$, is governed by the following second-order asymptotic expansion
	\begin{align}
	\label{eq:main_result}
	R(\rho,N,\epsilon) &\simeq D(\rho \| \G(\rho))+\frac{\Phi^{-1}(\epsilon)}{\sqrt{N}} \sqrt{V(\rho \| \G(\rho))},
	\end{align}
	with $\Phi^{-1}$ denoting the inverse function of the normal Gaussian cumulative distribution function $\Phi$. Moreover, the optimal rate is achieved using product unitary encodings belonging to $G^{\times N}$.
\end{restatable} 

\begin{proof}

In order to prove Theorem~\ref{thm:asymptotic}, we simply need to use Theorem~\ref{thm:single-shot} and the second order asymptotic expansions of the information spectrum relative entropy. First, it was established in~\cite{tomamichel12} (see also~\cite{li2014second}) that for every two density matrices $\rho,\sigma$, fixed $0<\epsilon<1$ and $\delta=O(1/\sqrt{N})$ we have
\begin{equation}
\!\!\! \frac{1}{N}D_{s}^{\epsilon \pm \delta}\left(\rho^{\otimes N} \| \sigma^{\otimes N}\right)\simeq D(\rho \| \sigma)+\frac{\Phi^{-1}(\epsilon)}{\sqrt{N}}\sqrt{V(\rho \| \sigma)}.
\end{equation} 
Now, we employ the definition of the encoding rate, Eq.~\eqref{eq:rate}. Substituting $\delta = 1/\sqrt{N}$ into the bounds in Theorem~\ref{thm:single-shot} and applying the above expansion, we obtain
\begin{align}
\label{eq:optimality_result}
R(\rho,N,\epsilon) &\simeq D(\rho \| \G(\rho))+\frac{\Phi^{-1}(\epsilon)}{\sqrt{N}} \sqrt{V(\rho \| \G(\rho))}.
\end{align} 

\end{proof}

\begin{rmk}
	Using the definition of a c-q channel from Remark~\ref{rmk:cq-channel}, we can interpret achievability of the rate in Eq.~\eqref{eq:main_result} for finite groups as an achievable rate for the corresponding c-q channel. In this case, the achievability thus follows from Refs.~\cite{schumacher97}~and~\cite{tomamicheltan14}.
\end{rmk}

\section{Applications}
\label{sec:discussion}

\subsection{Encoding power of unitary subgroups}
\label{sec:subgroups}

We will now discuss some consequences of the fact that the optimal encoding rate into resources destroyed by $\G^{\otimes N}$ can be achieved by local unitaries $U_{g_1}\otimes\dots\otimes U_{g_N}$, where each $U_{g_i}$ belongs to $G$. First, let us consider $G$ being the full unitary group $U$. Then, $\G^{\otimes N}$ is the completely depolarising channel, so encodings into resources destroyed by $\G^{\otimes N}$ coincide with general (i.e., non-local) unital encodings into $N$ systems. Employing Theorem~\ref{thm:asymptotic}, we then have the following.
\begin{cor}
	\label{cor:unitary}
	  Local unitary encodings achieve (up to second order asymptotic terms) the same optimal encoding rate $R_U(\rho,N,\epsilon)$ as global unital encodings:
	\begin{subequations}
		\begin{align}
			\label{eq:uni1}
			R_U(\rho,N,\epsilon)&\simeq R_{U}(\rho) + \frac{\Phi^{-1}(\epsilon)}{\sqrt{N}} \sqrt{V(\rho)},\\
			R_U(\rho)&:=\log d - S(\rho),\label{eq:uni2}
		\end{align}
	\end{subequations}
	where $S(\rho):=-\tr{\rho\log \rho}$ is the von Neumann entropy and \mbox{$V(\rho):=\tr{\rho(S(\rho)+\log\rho)^2}$} is the entropy variance.	
\end{cor}
\begin{rmk}
	The statement of Corollary~\ref{cor:unitary} still holds if the local unitaries $U_{g_i}$ do not belong to the full set of unitary matrices, but to an arbitrary unitary 1-design.
\end{rmk}
\noindent Moreover, as already mentioned in Sec.~\ref{sec:intro}, Eqs.~\eqref{eq:uni1}-\eqref{eq:uni2} not only specify the encoding power of local unitary transformations acting on $\rho^{\otimes N}$, but also the amount of information that can be encoded in the resource of purity of the state $\rho^{\otimes N}$. 

Now, if $G$ is a proper subgroup of the full unitary group, Theorem~\ref{thm:asymptotic} gives us that the optimal encoding rate $R_{\mathrm{G}}(\rho,N,\epsilon)$ into resources destroyed by the $G$-twirling channel $\G^{\otimes N}$ is given by
\begin{subequations}
	\begin{align}
	R_{\mathrm{G}}(\rho,N,\epsilon)&\simeq R_{\mathrm{G}}(\rho) + \frac{\Phi^{-1}(\epsilon)}{\sqrt{N}} \sqrt{V(\rho\|\G(\rho))},\label{eq:subuni1}\\
	R_{\mathrm{G}}(\rho)&:=D(\rho\|\G(\rho))=S(\G(\rho))-S(\rho).\label{eq:subuni2}
	\end{align}
\end{subequations}
Note that a more explicit expression for \mbox{$V(\rho\|\G(\rho))$} can be obtained by using the covariance
\begin{equation}
	\mathrm{cov}_{\!\rho}(A,B)= \tr{\rho A B}-\tr{\rho A}\tr{\rho B},
\end{equation}
which allows one to write
\begin{equation}
	V(\rho\|\G(\rho))=V(\rho)+V(\G(\rho))+2\mathrm{cov}_{\rho}(\log\rho,\log\G(\rho)).
\end{equation}
Combining Eqs.~\eqref{eq:uni1}-\eqref{eq:uni2} with \eqref{eq:subuni1}-\eqref{eq:subuni2}, one obtains the following corollary of Theorem~\ref{thm:asymptotic}.
\begin{cor}
	\label{cor:subunitary}
	 For any subgroup $G$ of unitary encodings, the optimal asymptotic encoding rate $R_U$ achieved by local unitary channels admits the following additive splitting 
	\begin{equation}
	\label{eq:splitting}
	R_{U}(\rho)=R_{G}(\rho)+R_{U}(\G(\rho)),
	\end{equation}
	where $R_G$ is the optimal asymptotic encoding rate achieved by local unitary channels belonging to $G$. 
\end{cor}

\begin{rmk}
	The statement of Corollary~\ref{cor:subunitary} still holds if the optimal asymptotic encoding rate $R_U$ achieved by local unitary channels is replaced by $R_{G'}$, the optimal asymptotic encoding rate achieved by local unitary channels belonging to $G' $, as long as $G\subset G'$.
\end{rmk}

Corollary~\ref{cor:subunitary} can be interpreted as follows. In the asymptotic limit, $N\rightarrow\infty$, the number of messages that can be encoded into $\rho^{\otimes N}$ per one copy of $\rho$ using all local unitary transformations splits additively into two terms. The first one corresponds to the optimal encoding rate when one is constrained to a subgroup $G$ of all unitary encodings. The second term tells us about the maximal number of messages that can be encoded using all unitaries, but on a state with resources erased by $\G$. In short: the full unitary encoding power for $\rho$ is just the sum of the encoding power of $G$ for $\rho$ and the full unitary encoding power for $\G(\rho)$.

Finally, let us note that the above considerations are closely related to the problem of distinguishing between unitary channels $U_{g}(\cdot)U_{g}^{\dagger}$ using an input state $\rho$, i.e., distinguishing between states \mbox{$\sigma_{g}:=U_{g}\rho U_{g}^{\dagger}$}. In this problem, the number of messages to be encoded is fixed (and equal to the order of the group), and one wants to maximise the success probability of correctly guessing $\sigma_{g}$. The authors of Ref.~\cite{piani2016robustness} showed that this success probability is directly related to a resource measure known as robustness of asymmetry.

\subsection{Encoding information in quantum coherence}
\label{sec:discussion_coh}

For a given distinguished orthonormal basis $\{\ket{k}\}_{k=1}^d$, the diagonal elements of $\rho$, $\matrixel{k}{\rho}{k}$, are known as populations, while the off-diagonal elements, $\matrixel{k}{\rho}{l}$ with $k\neq l$, are called coherences. The completely dephasing quantum channel $\Delta$ with respect to this basis sends all coherences to zero whilst not affecting the populations. More generally, we say that a quantum channel $\E$ is \emph{population-preserving} if for all $k,l$ we have
\begin{equation}
	\bra{k}\E(\ketbra{l}{l})\ket{k}=\delta_{kl},
\end{equation}
with $\delta_{kl}$ denoting the Kronecker delta. In other words, such channels process only coherences of a quantum system, and are also known in the literature as \emph{generalised dephasing channels}~\cite{devetak2005capacity}. 

As already noted in Sec.~\ref{sec:intro}, channels $\E_m$ that are encodings into resources destroyed by $\Delta$ are precisely the population-preserving channels. Using Theorems~\ref{thm:single-shot}~and~\ref{thm:asymptotic}, we can thus study the capacity of coherence to carry information. 

\begin{cor}
	\label{cor:diagonal}
	Local diagonal unitary encodings achieve (up to second order asymptotic terms) the same optimal encoding rate $R_\Delta(\rho,N,\epsilon)$ as global population-preserving encodings:
	\begin{subequations}
		\begin{align}
		\label{eq:rate_coh1}
		R_{\mathrm{\Delta}}(\rho,N,\epsilon)&\simeq R_{\mathrm{\Delta}}(\rho) + \frac{\Phi^{-1}(\epsilon)}{\sqrt{N}} \sqrt{V(\rho\|\Delta(\rho))},\\
		R_{\mathrm{\Delta}}(\rho)&:=D(\rho\|\Delta(\rho))=S(\Delta(\rho))-S(\rho).	\label{eq:rate_coh2}
		\end{align}
	\end{subequations}	
\end{cor}
\begin{rmk}
	The statement of Corollary~\ref{cor:diagonal} still holds if the local unitaries $U_{g_i}$ do not belong to the full set of diagonal unitary matrices, but to an arbitrary subset the twirling over which yields the completely dephasing channel. Examples of such subsets include a set of $d$ diagonal matrices with diagonals given by the rows of the $d$-dimensional Fourier matrix, and a set of $2^d$ diagonal matrices whose $j$-th diagonal elements are given by $(-1)^{x_j}$, with $\v{x}$ denoting a binary vector of length $d$.
\end{rmk}

Corollary~\ref{cor:diagonal} provides operational meaning to measures of coherence studied within the resource theory of coherence~\cite{baumgratz2014quantifying} by relating them to the optimal rate of encoding classical information into coherences. In particular, $D(\rho\|\Delta(\rho))$ appearing in Eq.~\eqref{eq:rate_coh2} is the well-known relative entropy of coherence that quantifies distillable coherence in the asymptotic limit under incoherent operations~\cite{winter2016operational} (as well as distillable coherence and coherence cost within the framework of dephasing-covariant incoherent operations~\cite{chitambar2018dephasing}); while $V(\rho\|\Delta(\rho))$ is the relative entropy variance of coherence which, to the authors' best knowledge, is first introduced here. Moreover, it is clear from Corollary~\ref{cor:diagonal} that the states that are asymptotically optimal for encoding information into coherences are pure states (so that $S(\rho)=0$) that get dephased to a maximally mixed state (so that $S(\Delta(\rho))=\log d$ is maximal). The general form of such a state is given by a uniform superposition of all states from the distinguished basis. Finally, using the additive splitting from Corollary~\ref{cor:subunitary}, we also have that the amount of information that can be unitarily encoded in states $\rho^{\otimes N}$ is equal to the amount of information that can be encoded in coherences of $\rho^{\otimes N}$ plus the amount of information that can be encoded in a decohered state $\Delta(\rho)^{\otimes N}$. Note that this splitting is directly related to decomposing uncertainty into classical and quantum parts~\cite{korzekwa2014quantum}. 

Let us also point out that the problem of encoding information into coherences was studied before in the single-shot and error-free scenario. First, in Ref.~\cite{korzekwa2018coherifying}, the concept of coherifying quantum states was introduced: a state $\rho$ is a coherification of a diagonal state $\rho_0$ if it dephases to it, i.e., $\D(\rho)=\rho_0$. Then, in Ref.~\cite{korzekwa2019distinguishing}, the authors were investigating the number of coherifications of $\rho_0$ with non-overlapping supports or, in other words, the number of orthogonal states that are classically indistinguishable (as they are sent to the same state by a dephasing map $\Delta$). The number of such perfectly distinguishable states was also related to time-energy uncertainty relation~\cite{coles2019entropic}. We can thus employ the current results to study a particular case of time-energy uncertainty relation for the clock system that is composed of many independent and identically distributed quantum states. More precisely, and specialising to the asymptotic scenario captured by Theorem~\ref{thm:asymptotic}, consider $N$ copies of a quantum system, each described by a Hamiltonian $H=\sum_k E_k\ketbra{E_k}{E_k}$ and prepared in a pure state $\ket{\psi}$. Then, the ability of $\ket{\psi}^{\otimes N}$ to act as a clock can be measured by the number $T$ of states distinguishable with failure probability $\epsilon$ that it passes through during a free evolution generated by the total Hamiltonian~\cite{korzekwa2019distinguishing}. From Eqs.~\eqref{eq:rate_coh1}-\eqref{eq:rate_coh2}, we see that asymptotically for vanishing error $\epsilon$ this number is upper bounded as $T\leq 2^{h(\v{p})N}$, where $h(\v{p})=-\sum_k p_k\log p_k$ is the Shannon entropy of the energy distribution $p_k:=|\braket{E_k}{\psi}|^2$. Therefore, the better resolution of the clock we want to get, the higher entropy of energy distribution is required, and so the inequality
\begin{align}
	\frac{1}{\log T} \cdot h(\v{p}^{\otimes N}) \geq 1,
\end{align}
can be interpreted as time-energy uncertainty relation. Of course, one could also use the second order term from Eq.~\eqref{eq:rate_coh1} to get an even tighter result.

\subsection{Asymptotic super-dense coding}

Consider now encoding information into a state $\rho_{AB}$ of a bipartite system $AB$ (with local dimensions $d_A$ and $d_B$), using encodings into resources destroyed by the $G$-twirling channel $\G$ over all unitaries on $A$. As we show in Appendix~\ref{app:local_unital}, such encodings correspond to local unital channels on system $A$. Using Theorem~\ref{thm:asymptotic}, we can then find the number of approximately orthogonal states that the global state of $AB$ can be steered to by operating only locally on $A$. In other words, we can study the optimal rate of unital super-dense coding, which is captured by the following corollary.

\begin{cor}
	\label{cor:superdense}
	Product unitary encodings on subsystem $A$ achieve (up to second order asymptotic terms) the same optimal super-dense coding rate $R_{\mathrm{loc}}(\rho_{AB},N,\epsilon)$ as unital encodings on subsystem $A$:
	\begin{subequations}
		\begin{align}
		R_{\mathrm{loc}}(\rho_{AB},N,\epsilon)&\simeq R_{\mathrm{loc}}(\rho_{AB}) +  \frac{\Phi^{-1}(\epsilon)}{\sqrt{N}} \sqrt{V(A|B)}  \,,\\
		R_{\mathrm{loc}}(\rho_{AB})&:=\log d_A - S(A|B),
		\end{align}
	\end{subequations}
	with $S(A|B) = -D(\rho_{AB} \| \iden_A \otimes \rho_B) = S(\rho_{AB})-S(\rho_B)$ being the conditional entropy, and \mbox{$V(A|B) = V(\rho_{AB} \| \iden_A \otimes \rho_B)$} being the corresponding variance~\cite{tomamichel12}.
\end{cor}

In the simplest case of a two-qubit system, Corollary~\ref{cor:superdense} recovers the asymptotic rate for super-dense coding~\cite{bennett1992communication}: if $\rho_{AB}$ is a pure product state, we can encode only a single bit per a copy of $\rho_{AB}$ by operating on $A$; but if $\rho_{AB}$ is one of the four Bell states, we can encode 2 bits per copy, since local operations can map between all Bell states. More generally, the optimal encoding rate from Corollary~\ref{cor:superdense} recovers the known asymptotic result obtained first for qubit systems in Ref.~\cite{bowen2001classical}, and then generalised to qudits in Refs.~\cite{hiroshima2001optimal,horodecki2001classical}, but also yields the second order asymptotic correction term. As such, it complements the results of Ref.~\cite{datta2014second}, where the second order corrections for super-dense coding were derived under the simplifying assumption of product encodings (but without assuming unitality). Since the obtained expressions are identical, our result proves that no amount of correlated encodings into many copies of the system will improve the encoding rate if one is limited to unital channels. Also, using again the splitting from Eq.~\eqref{eq:splitting}, we find that the amount of information that can be encoded via global unitaries in $N$ copies of a state $\rho_{AB}$ is equal to the amount of information that can be encoded in $\rho_{AB}$ via local unitaries on $A$ plus the amount of information that can be encoded unitarily in $\rho_B$:
\begin{equation}
	\label{eq:splitting_super}
	R_{U}(\rho_{AB})=R_{\mathrm{loc}}(\rho_{AB})+R_{U}(\rho_B).
\end{equation}
Finally, it is clear from Corollary~\ref{cor:superdense} that the states that are asymptotically optimal for super-dense coding are pure states with a maximally mixed marginal, i.e., maximally entangled states.

\subsection{Collective encoding and encoding via permutations}
\label{sec:collective}

Let us now switch to an $n$-partite system with equal Hilbert spaces of dimension $d$, prepared in a state $\rho_{1\dots n}$. We want to discuss encodings into resources destroyed by two types of resource destroying maps, $\G_{\mathrm{col}}$ and $\G_{\mathrm{per}}$, acting on each $n$-partite system via
\begin{subequations}
	\begin{align}
	\G_{\mathrm{col}} (\cdot) &= \int_{\mathrm{U(d)}} dg ~ U(g)^{\otimes n}(\cdot)U^\dagger(g)^{\otimes n},\label{eq:G_col}\\
	\G_{\mathrm{per}} (\cdot) &= \frac{1}{n!}\sum_{\pi_i\in S_n} \pi_i(\cdot)\pi_i^\dagger\label{eq:G_per}.
	\end{align}
\end{subequations}
Here, the integral in the first equation goes over all $d$-dimensional unitaries $U(g)$ (according to the Haar measure), and the sum in the second equation is over all permutations $\pi_i$ between $n$ subsystems. In the case of $\G_{\mathrm{col}}$, Theorems~\ref{thm:single-shot}~and~\ref{thm:asymptotic} allow us to study the encoding power of collective unitaries, i.e., the amount of information that can be encoded into $\rho_{1\dots n}$ (or $N$ copies of it) when the allowed operations are given by the same unitary on each subsystem. This can be interpreted as encoding information in the global degrees of freedom, as such unitaries cannot affect relative degrees of freedom between different subsystems. And, in the case of $\G_{\mathrm{per}}$, we can investigate how much information can be encoded by just permuting subsystems between $n$ parties.

In order to find the optimal number of messages that can be encoded, we need to understand how the twirling operations $\G_{\mathrm{col}}$ and $\G_{\mathrm{per}}$ act on a general state $\rho_{1 \dots n}$. In the simplest case of a bipartite system, $n=2$, their action takes a particularly simple form:
\begin{subequations}
\begin{align}
	\G_{\mathrm{col}}(\rho_{12})&=p_s\frac{\Pi_s}{d_s}+(1-p_s)\frac{\iden-\Pi_s}{d^2-d_s},\\
	\G_{\mathrm{per}}(\rho_{12})&=\Pi_s \rho_{12} \Pi_s+(\iden-\Pi_s) \rho_{12} (\iden-\Pi_s),
\end{align}
\end{subequations}
where $p_s=\tr{\rho_{12}\Pi_s}$, $d_s=\tr{\Pi_s}=d(d+1)/2$ and $\Pi_s$ is the projector onto the symmetric subspace, i.e.,
\begin{equation}
\Pi_s=\sum_{k=1}^d \ketbra{kk}{kk} + \sum_{k=1}^d\sum_{l=k+1}^d\!\! \frac{(\ket{kl}\!+\!\ket{lk})(\bra{kl}\!+\!\bra{lk})}{2}.
\end{equation}
We can now ask, which states $\rho_{12}$ are asymptotically optimal for encoding information using collective unitaries, and which ones are optimal for encoding information using permutations. In the first case, these are given by pure states
\begin{align}
	\ket{\psi^*_{12}}=\sqrt{\frac{d+1}{2d}}\ket{\psi_s} + \sqrt{\frac{d-1}{2d}}\ket{\psi_a},
\end{align}
where $\ket{\psi_s}$ and $\ket{\psi_a}$ are arbitrary pure states living in the symmetric and antisymmetric subspaces, respectively. Such states are optimal, as they are pure and twirl to a maximally mixed state. In the case of encoding information using permutations, the situation is even simpler, as there are only two allowed channels (identity and transposition between the two systems). Thus, the maximal number of messages we can encode per one copy of the system is upper bounded by 2. This bound can be attained, even in single-shot scenario, by simply choosing the state $\ket{01}$.

Beyond the above bipartite examples, one could further investigate multipartite scenarios for $n>2$. In order to find the action of $\G_{\mathrm{col}}$ and $\G_{\mathrm{per}}$ (and so to derive the optimal encoding rates), one could employ the fact that these two resource destroying maps are closely related via a Schur-Weyl duality~\cite{marvianthesis}. That is, the permutation group and the group of collective unitaries commute, and so the tensor space decomposes into a direct sum of tensor products of irreducible modules for these two groups. In particular, for $\G_{\mathrm{col}}$ we could ask whether there always exists a quantum state $\rho_{1 \dots n}$ that under collective unitaries can asymptotically encode the ultimate maximal number of bits per system (which is specified by the dimensionality of $\rho_{1 \dots n}$, i.e., $\log d^n$). Thus, the problem is to find a pure state $\ket{\psi}$ that is mapped to a maximally mixed state by $\G_{\mathrm{col}}$. In Appendix~\ref{app:collective} we briefly discuss how to approach this problem and provide an example for a tripartite system. In the case of $\G_{\mathrm{per}}$ we could instead ask, whether it is always possible to find a state $\rho_{1 \dots n}$ that maps under permutations to a mutually orthogonal set of states, thus encoding the maximal number of messages \mbox{$\min\{n!,d^n\}$}. For example, it is easy to see that when $d\geq n$, one can simply choose a state \mbox{$\ket{\psi_\pi}=\ket{1,2,\dots,d}$}, which is mapped by permutations to $n!$ orthogonal states, and so it is optimal for encoding messages with permutation group. On the other hand, when $d<n$, a plausible ansatz for the optimal state is given by \mbox{$\ket{\psi_\pi}^{\otimes \lfloor \frac{n}{d}\rfloor} \otimes \ket{\phi_\pi}$} with $\ket{\phi_\pi}=\ket{1,2,\dots,n-d\lfloor \frac{n}{d}\rfloor}$.

\subsection{Shared reference frames and private communication}

We now want to briefly discuss the relation between encoding into resources destroyed by $\G$ and a private classical communication scheme for a decohering superoperator $\G$~\cite{bartlett2004decoherence}, as introduced in the studies on quantum reference frames~\cite{bartlett2007reference}. In this scenario there are three parties: beyond the sender $S$ and the receiver $R$, there is also an eavesdropper $E$. The two communicating parties share a private reference frame for some degree of freedom, i.e., both $S$ and $R$ agree on the form of group representation $U(g)$ given a classical label $g$ describing it. As an example, consider a shared Cartesian frame of reference (given, e.g., by three mutually orthogonal rigid rods defining directions $x,y,z$). Then, the classical description of an element of the rotation group can be given by three Euler angles with respect to the axes defined by the shared reference frame. Thus, if $S$ tells $R$ that she prepared a spin-1/2 particle along the positive $z$ direction, and he wants to rotate it, so that it points in the opposite direction, he knows precisely which $U(g)$ to perform. However, $E$ may not have full access to that reference frame. Let us consider here two cases: in the first one, she does not have any information about the orientation of the reference frame; and in the second one, she knows the orientation of the $z$ axis, but does not know how $x$ and $y$ are oriented. Therefore, in the first case, her description of the system is given by a uniform mixture (stemming from no knowledge about the orientation of the reference frame) over all possible rotations of the reference frame. This way a state $\rho$ is described by $E$ as $\G(\rho)$, where $\G$ is the twirling over SO(3) group, and so any initial state of a spin-1/2 particle looks to $E$ like a maximally mixed state. In the second case, her description of the system is given by a uniform mixture over all possible rotations in the $xy$ plane. This way a state $\rho$ is described by $E$ as $\G(\rho)$, where $\G$ is the twirling over SO(2) group, and so any initial state $\rho$ of a spin-1/2 particle looks to $E$ like a dephased version of $\rho$, with dephasing in the eigenbasis of spin-up/spin-down states along the $z$ direction. Of course, in a general case the twirling can happen over arbitrary group $G$ corresponding to a shared reference frame between $S$ and $R$ that $E$ has no access to. 

Now, following the definition given in Ref.~\cite{bartlett2004decoherence}, $S$ and $R$ have a private classical communication scheme employing a shared reference frame related to a group $G$, if $S$ can prepare $M$ orthogonal states $\sigma_m$, such that $\G(\sigma_m)=\rho_0$ for all $m$ and some fixed state $\rho_0$. This means that $S$ can send one of $M$ perfectly distinguishable messages to $R$, while at the same time for the eavesdropper $E$ all these messages will be completely indistinguishable, and so the communication will be secure. Through Eq.~\eqref{eq:constr_2}, it is then clear that if for a state $\rho$ there exists $M$ encodings into resources destroyed by $\G$, then $S$ and $R$ have a private classical communication scheme: $S$ prepares one of the states $\E_m(\rho)$ that are (almost) perfectly distinguishable by $R$, but for $E$ they are all described by $\G(\E_m(\rho))=\G(\rho)$. Thus, each state $\rho$ specifies a private communication scheme allowing for the secure communication of $\log M$ bits, and our Theorem~\ref{thm:single-shot} provides upper and lower bounds for $\log M$ in the single-shot regime. Optimising the information spectrum relative entropy between $\rho$ and $\G(\rho)$ over all states $\rho$ then yields the optimal scheme for secure communication.

Note however, that for the asymptotic result from Theorem~\ref{thm:asymptotic} to hold for private communication using $\rho^{\otimes N}$, $S$ and $R$ require many copies of uncorrelated reference frames -- otherwise $E$ can learn the orientation of the reference frame from the first few copies of $\rho$. Alternatively, one could use the setup of $\G_{\rm col}$: namely, instead of performing the asymptotic analysis as in Theorem~\ref{thm:asymptotic}, one could use the one-shot bounds and increase the number of parties that $\G_{\rm col}$ twirls over (i.e.\ we go $n \to \infty$ keeping $N = 1$). More precisely, instead of using $N$ independent reference frames (each for one copy of $\rho$) that leads to $E$ describing the encoded message as $\G(\rho)^{\otimes N}$, one could use a single reference frame for all $n$ states, so that $E$ would see the encoded message as $\G_{\rm col}(\rho^{\otimes n})$. The general procedure of performing such a collective twirling can be found, e.g., in Ref.~\cite{bartlett2007reference}, and here we will just present a simple example illustrating that, even with a single reference frame, increasing the number $n$ of identical states can increase the number of securely communicated messages. Consider $A$ and $B$ sharing a Cartesian reference frame (that $E$ has no access to) and trying to communicate using $n$ spin-1/2 systems, with each spin initially prepared in the same state $\ket{\uparrow}$ along the $z$ axis. Now, collective $G$-twirling of a state $\rho$ of $n$ spin-1/2 systems first projects it onto subspaces with total angular momentum $j\in\{n/2-\lfloor n/2 \rfloor,\dots,n/2\}$, and then completely depolarises it within the irreducible subspaces, not affecting the multiplicity subspaces (see Ref.~\cite{bartlett2007reference} for details). Since the state $\ket{\uparrow}^{\otimes n}$ belongs to the subspace with maximal angular momentum, $j_{\max}=n/2$, and because this subspace does not have any multiplicities, the twirled state is $\G(\ketbra{\uparrow}{\uparrow}^{\otimes n})=\iden_{j_{\max}}/(2j_{\max} +1)$, i.e. the maximally mixed state over the $2j_{\max}+1$ states within the $j_{\max}$ subspace. As a result, the number of securely communicated bits from $A$ to $B$, $\log M$, with decoding error $\epsilon\leq 1/2$ is bounded by Theorem~\ref{thm:single-shot} as
\begin{equation}
	\log (n+1)  - \log\frac{2}{\epsilon}\leq \log M \leq \log (n+1)  + \log\frac{1}{\epsilon}.
\end{equation}	
Note that in this simple example one can easily find a code that yields $M=n+1$ with zero decoding error: one simply needs to encode messages in $2j_{\max}+1$ orthogonal states $\{\ket{j_{\max},m}\}_{m=-j_{\max}}^{j_{\max}}$ that all look like a maximally mixed state to $E$.

\subsection{Thermodynamics}

Finally, we want to make a short comment on the case of a resource destroying map given by the completely thermalising map $\T$. Since it is not a $G$-twirling channel, we cannot use Theorems~\ref{thm:single-shot}~or~\ref{thm:asymptotic}, but Lemma~\ref{lemma:upper} still applies in this case. Thus, in the asymptotic limit, the number $M$ of almost orthogonal states (i.e., distinguishable with probability $1-\epsilon$) that can be obtained via Gibbs-preserving operations (that model free thermodynamic transformations with no external access to any sources of work or sinks of entropy) from a state $\rho^{\otimes N}$ is upper bounded as
\begin{equation}
	\frac{\log M(\rho^{\otimes N},\epsilon)}{N}\lesssim  D(\rho \| \gamma)+\frac{\Phi^{-1}(\epsilon)}{\sqrt{N}}\sqrt{V(\rho \| \gamma)}.
\end{equation}
Now, the crucial thing is that the entropic quantities appearing on the right hand side of the above inequality have a clear thermodynamic interpretation. Denoting by $\beta$ the inverse temperature, we have that $D(\rho \| \gamma)/\beta$ is the free energy of the state~$\rho$ that quantifies the amount of useful work which can be extracted from asymptotically many copies of $\rho$ when using only free thermodynamic transformations~\cite{brandao2013resource}. Similarly, $V(\rho \| \gamma)/\beta$ can be interpreted as the free energy fluctuations of the state~$\rho$~\cite{biswas2021fluctuation}, and was recently shown to quantify the amount of work dissipated due to finite number of copies of $\rho$ (i.e., it describes the second order correction to the asymptotic result given by free energy) for the case when $\rho$ is classical~\cite{chubb2017beyond} or pure \cite{biswas2021fluctuation}. We thus see that the number of messages that can be thermodynamically encoded for free in a state $\rho^{\otimes N}$ is bounded by the quantity directly related to the amount of thermodynamic work that one can perform while thermalising~$\rho^{\otimes N}$. We also note that the authors of Ref.~\cite{narasimhachar2019quantifying} have recently made a similar observation while quantifying memory capacity as a thermodynamic resource within the framework of thermal operations. Putting it another way, the volume of the future thermal cone of $\rho$, i.e., the number of distinguishable states it can evolve to during the evolution generated by interaction with the heat bath, is upper bounded by the amount of work one can extract from~$\rho$.

\section{Outlook}
\label{sec:outlook}

In this paper we have studied the problem of single-shot and asymptotic encoding of classical information in resources of a quantum state $\rho$ destroyed by a decohering quantum channel $\D$, i.e., in the degrees of freedom that completely decohere under the action of $\D$. We focused on a particular family of resource destroying maps given by $G$-twirling operators $\G$ over arbitrary unitary subgroups. In Theorem~\ref{thm:single-shot} we found lower and upper bounds for the number of messages that can be encoded in resources destroyed by $\G$ with an error probability $\epsilon$; while in Theorem~\ref{thm:asymptotic} we found the second order asymptotic expansion for the encoding rate. We then discussed applications of our results to a number of problems in quantum information theory, including quantifying informational capacity of quantum coherence, usefulness of entangled states for super-dense coding and encoding power of unitary subgroups.

We believe that an interesting area of research concerns studying more realistic settings, when the transmission channel between the sender and receiver is not perfect, but noisy. Although this topic deserves a more detailed study, we can already make some basic observations here. For a resource destroying map given by the $G$-twirling operator $\G$, one can easily accommodate for the noise $\N_\G^\lambda$ given by a partial $G$-twirling operator: 
	\begin{equation}
	\mathcal{N}_\G^\lambda(\rho)=(1-\lambda)\rho + \lambda \G(\rho).
	\end{equation}
Due to the constraints on encoding maps $\E_m$ given in Eqs.~\eqref{eq:constr_1}-\eqref{eq:constr_2}, they will commute with the noise. This means that the set of encoded messages $\{\E_m(\rho)\}$ after the transmission to the receiver will be replaced by $\{\E_m(\N_\G^\lambda(\rho))\}$. Thus, the effect of a noisy communication channel will simply be captured by applying the noise to the initial state of the information carrier, i.e., the optimal single-shot and asymptotic encoding rates will still be described by Theorems~\ref{thm:single-shot}~and~\ref{thm:asymptotic}, but with $\rho$ replaced by $\N_\G^\lambda(\rho)$. For example, the asymptotic encoding rate into coherences of $\rho$, analysed in Sec.~\ref{sec:discussion_coh}, will decrease from $D(\rho\|\Delta(\rho))$ to $D(\N^\lambda_\Delta(\rho)\|\Delta(\rho))$ due to a partially dephasing channel $\N^\lambda_\Delta$ between the sender and receiver. Moreover, when $\D$ is the $G$-twirling operator, encodings $\E_m$ are constrained by Eqs.~\eqref{eq:constr_1}-\eqref{eq:constr_2} to be unital. This means that encodings into resources destroyed by $\G$ for an arbitrary group $G$ will commute with a partially depolarising channel,
	\begin{equation}
		\mathcal{N}^\lambda(\rho)=(1-\lambda)\rho + \lambda \frac{\iden}{d}.
	\end{equation}
	By the same argument as before, the only effect induced by a partially depolarising channel $\N^\lambda$ between the sender and receiver will be captured by replacing $\rho$ with $\N^\lambda(\rho)$ in Theorems~\ref{thm:single-shot}~and~\ref{thm:asymptotic}.

Another interesting line of research, that we have only touched in Sec.~\ref{sec:subgroups}, concerns scenarios with multiple complementary resources. We have seen here that the encoding power of a full unitary group can be additively decomposed into the encoding power of a given subgroup $G$, and the encoding power of a full unitary group for a $G$-twirled state. This shows that the amount of ``storage space'' left after using the asymmetry resource of a state is given by the remaining resource of purity, meaning that these two resources behave in a complementary way (the more asymmetry, the less purity in a $G$-twirled state). More generally, one can consider encoding of information into two distinct resources destroyed by $\D_1$ and $\D_2$. The communication scenario one could envisage then is that with two senders, $S_1$ and $S_2$, and one receiver $R$. First, $S_1$ encodes one of $M_1$ messages into resources destroyed by $\D_1$, sends the encoded state to $S_2$, who encodes one of $M_2$ messages into resources destroyed by $\D_2$ on top of the previous one, and finally sends it to $R$. In such a scenario it is then natural to investigate when the receiver can decode both messages, and what is the optimal trade-off between $M_1$ and $M_2$. For example, considering encoding into coherences with respect to two distinct bases, one would expect to see a trade-off related to non-commutativity and the resulting uncertainty relations; or, while encoding into asymmetry relative to subsystem permutations and collective unitaries, as discussed in Sec.~\ref{sec:collective}, the Schur--Weyl duality may result in these resources behaving in a complementary way. This multi-resource considerations, however, require different proof techniques than the ones used in the current paper and surely deserve a separate study.

There are also several other clear paths for future research stemming from our results. First, one could look for other unitary subgroups with operational relevance, and thus find second-order asymptotic encoding rates for constrained communication scenarios. Second, we expect that not only upper bound holds for general resource destroying maps $\D$ (Lemma~\ref{lemma:upper}), but also that there should be a lower bound that asymptotically coincides with the upper one. In particular, it would be interesting to prove the existence of such bounds for $\D$ being given by the completely thermalising map $\T$. This way one would relate the encoding power of Gibbs-preserving operations in a state $\rho$ with the amount of work that can be extracted from $\rho$ with these operations, thus providing one more strong link between information theory and thermodynamics. Finally, as the optimal encoding rate in resources destroyed by $\G$ is given by $S(\G(\rho))-S(\rho)$, one could look for states that maximise this quantity for general groups $G$. More broadly, one could also investigate the generalisation of the concept of coherification, with \emph{asymmetrization} of a symmetric state $\rho_0=\G(\rho_0)$ being given by any state $\rho$ such that $\G(\rho)=\rho_0$. In particular, the number of orthogonal asymmetrizations of $\rho$ would be then directly related to the number of classical messages that one can encode in the resources destroyed by $\G$.

\textbf{Acknowledgements:} The authors would like to thank the anonymous referee of our previous work~\cite{korzekwa2019distinguishing}, who suggested an asymptotic analysis of the number of orthogonal states with the same diagonal and inspired us to start this project. We are also indebted to the final referee of this work, who wrote an exceptionally detailed and constructive report on our paper and allowed us to improve it considerably. Furthermore, we are grateful to Christopher T. Chubb for his insightful comments on local unital encodings, to Bartosz Regu{\l}a for useful comments concerning relative entropy of coherence, and to Varun Narasimhachar for pointing us to a related work on thermodynamic encoding. KK would also like to thank D.~Jennings and C. C\^{i}rstoiu for helpful discussions. We acknowledge financial support by the Foundation for Polish Science through IRAP project co-financed by EU within Smart Growth Operational Programme (contract no. 2018/MAB/5) and through TEAM-NET project (contract no. POIR.04.04.00-00-17C1/18-00). K\.{Z} is supported by National Science Center in Poland under the Maestro grant number DEC-2015/18/A/ST2/00274. ZP is supported by the Polish National Science Centre under grant number 2016/22/E/ST6/00062. MT is supported in part by NUS startup grants (R-263-000-E32-133 and R-263-000-E32-731) as well as by the National Research Foundation,	Prime Minister’s Office, Singapore and the Ministry of Education, Singapore under the Research Centres of Excellence programme.

\appendix

\subsection{Group covariant encodings}

\label{app:covariant}

Despite some similarities, encodings into resources destroyed by $\G$ are distinct from the better known class of group covariant encodings~\cite{holevo1979covariant}. Let us recall that for a subgroup $G$ of unitary channels, $\{U_g(\cdot)U_g^{\dagger}\}_{g\in G}$, a group covariant quantum channel $\E$ satisfies
\begin{equation}
	\forall~g:~~ U_g\E(\cdot)U_g^{\dagger}=\E(U_g(\cdot)U_g^{\dagger}),
\end{equation}
whereas a channel $\E$ that is an encoding into resources destroyed by $\G$ must satisfy Eqs.~\eqref{eq:constr_1}-\eqref{eq:constr_2} that we repeat here for convenience:
\begin{equation}
	\label{eq:enco}
	\E(\G(\cdot))=\G(\E(\cdot))=\G(\cdot),	
\end{equation} 
We will now show that these two sets of channels are not comparable, i.e. neither one is a subset of the other one. 

First, consider any non-Abelian group $G$ and take a channel $\E$ to be given by
\begin{equation}
	\label{eq:ex1}
	\E(\cdot)=U_{h}(\cdot)U_h^{\dagger}
\end{equation}
for $h\in G$ that is not in the centralizer of $G$ (i.e. there exists at least one $g\in G$ such that $hg\neq gh$). It is then straightforward to see that such a channel is not group covariant. On the other hand, such a map satisfies Eq.~\eqref{eq:enco}, because
\begin{subequations}
\begin{align}
\E(\G(\cdot))&=\frac{1}{|G|}\sum_{g\in G}U_{h}U_{g}(\cdot)U_g^{\dagger}U_h^{\dagger}\nonumber\\
&=\frac{1}{|G|}\sum_{g\in G}U_{hg}(\cdot)U_{hg}^{\dagger}=\G(\cdot),\\
\G(\E(\cdot))&=\frac{1}{|G|}\sum_{g\in G}U_gU_h(\cdot)U_h^{\dagger}U_g^{\dagger}\nonumber\\
&=\frac{1}{|G|}\sum_{g\in G}U_{gh}(\cdot)U_{gh}^{\dagger}=\G(\cdot),
\end{align}
\end{subequations}
where in the last step we used the fact that \mbox{$\forall h\in G:~hG=Gh=G$}. Thus, $\E$ from Eq.~\eqref{eq:ex1} is not group covariant, but it is an encoding into resources destroyed by $\G$. 

Second, let us consider a subgroup $G$ with unitary matrices $U_{g}$ given by
\begin{equation}
	U_{g} = e^{iHx(g)},\quad H=\sum_{n=1}^d n\ketbra{n}{n},\quad x(g)\in[0,2\pi).
\end{equation}
Consider a quantum channel
\begin{equation}
\label{eq:ex2}
\E(\cdot)=\ketbra{1}{1}\tr{\cdot},
\end{equation}
which is obviously group covariant since
\begin{align}
\forall~g:~~\E(U_{g}(\cdot)U_g^{\dagger})=\ketbra{1}{1}&=U_g\ketbra{1}{1}U_g^{\dagger}\nonumber\\
&=U_g\E(\cdot)U_g^{\dagger}.
\end{align}
On the other hand, the considered channel is not an encoding into resources destroyed by $\G$, because we have
\begin{equation}
\G(\cdot)=\sum_{n=1}^d \bra{n}(\cdot)\ket{n} \ketbra{n}{n},
\end{equation}
and
\begin{subequations}
\begin{align}
\E(\G(\cdot))&=\ketbra{1}{1}\neq \G(\cdot),\\
\G(\E(\cdot))&=\G(\ketbra{1}{1})=\ketbra{1}{1}\neq \G(\cdot).
\end{align}	
\end{subequations}
Thus $\E$ from Eq.~\eqref{eq:ex2} is group covariant, but it is not an encoding into resources destroyed by $\G$.

\subsection{Local unital encodings}
\label{app:local_unital}

Conditions from Eq.~\eqref{eq:constr_1}-\eqref{eq:constr_2} for encodings $\E$ into resources destroyed by the $G$-twirling channel $\G$ over all unitaries on the $A$ subsystem of the composite $AB$ system read
\begin{subequations}
	\begin{align}
		\E\left(\frac{\iden_A}{d_A} \otimes \rho_B\right) &=  \frac{\iden_A}{d_A} \otimes \rho_B,\label{eq:app1}\\
		\mathrm{Tr}_A(\E(\rho_{AB})) &=  \rho_B,\label{eq:app2}
	\end{align}
\end{subequations}
where the above has to hold for all bipartite states $\rho_{AB}$, with $\rho_B$ denoting the reduced state on the $B$ subsystem.

We start from Eq.~\eqref{eq:app2}. Since it has to be satisfied for all input states $\rho_{AB}$, in particular it has to be satisfied for product states $\rho_{AB}=\rho_A\otimes \psi_B$ with $\psi_B$ being pure. We thus get
\begin{equation}
	\mathrm{Tr}_A(\E(\rho_{A}\otimes \psi_B)) =  \psi_B.
\end{equation}
However, we know that the only way a reduced state is a pure state is when we deal with a product state, so
\begin{equation}
	\label{eq:app_prod}
	\E(\rho_{A}\otimes \psi_B) =  \sigma_A\otimes \psi_B.
\end{equation}

Now, consider an arbitrary state $\rho_A$, and two arbitrary \emph{distinct} pure states $\psi_B$ and $\psi_B'$. From Eq.~\eqref{eq:app_prod}, there must exist states $\sigma_A$ and $\sigma_A'$ such that
\begin{subequations}
	\begin{align}
	\mathcal E(\rho_A\otimes\psi_B)=\sigma_A\otimes \psi_B,\label{eq:appb1}\\
	\mathcal E(\rho_A\otimes\psi_B')=\sigma_A'\otimes \psi_B'\label{eq:appb2}.
	\end{align}
\end{subequations}
Importantly, we note that we cannot assume \emph{a priori} that the output states $\sigma_A$ and $\sigma_A'$ will be independent of the state of the $B$ subsystem. We will now show, however, that the linearity of $\mathcal E$ requires that this is the case.

Consider the action of $\mathcal E$ on an arbitrary mixture of these two input states. Employing Eq.~\eqref{eq:app_prod} once more, this means that for any $t\in[0,1]$ there must exist a $\sigma''_A$ such that
\begin{align}
\label{eqn:lincomb}
\!\! \mathcal E\Bigl(\!\rho_A\otimes\!\bigl(t\psi_B+(1\!-\!t)\psi_B'\bigr)\!\Bigr)=\sigma_A''\otimes\!\bigl(t\psi_B+(1\!-\!t)\psi_B'\bigr).\!
\end{align}
Combining the above three equations with the linearity of $\mathcal E$ therefore yields
\begin{align}
\!\!\!t\sigma_A\otimes\! \psi_B+(1\!-\!t)\sigma_A'\otimes \psi_B'
=\sigma_A''\otimes\!\bigl(t\psi_B+(1\!-\!t)\psi_B'\bigr).\!\label{eq:appb3}
\end{align}
Taking a partial trace over the $B$ subsystem, this in turn tells us that the output on the $A$ subsystem must also simply be a mixture,
\begin{align}
\sigma_A''=
t\sigma_A+(1-t)\sigma_A'.\label{eq:appb4}
\end{align}
Now, calculating the left hand side of Eq.~\eqref{eqn:lincomb} once using Eqs.~\eqref{eq:appb1}-\eqref{eq:appb2}, and once using Eqs.~\eqref{eq:appb3}-\eqref{eq:appb4}, and equating the two expressions yields
\begin{align}
(\sigma_A-\sigma_A')\otimes (\psi_B-\psi_B')=0.
\end{align}
As we assumed that $\psi_B$ and $\psi_B'$ were distinct, we conclude that $\sigma_A=\sigma_A'$. As this argument holds for any $\rho_A$, we can therefore conclude that the output state $\sigma_A$ must indeed be independent of the subsystem $B$, and therefore that there exists some channel $\mathcal E_A$ such that
\begin{align}
	\mathcal E(\rho_A\otimes \psi_B)=\mathcal E_A(\rho_A)\otimes \mathcal \psi_B.
\end{align}
As this holds for all $\rho_A$ and $\psi_B$, we get that 
\begin{align}
	\label{eq:prod}
	\E=\E_A\otimes \mathcal I,
\end{align}
i.e., that $\E$ acts only locally on subsystem $A$.

Finally, combining Eq.~\eqref{eq:prod} with Eq.~\eqref{eq:app1}, we get
\begin{equation}
	\E\left(\frac{\iden_A}{d_A} \otimes \rho_B\right) = \E_A\left(\frac{\iden_A}{d_A}\right)\otimes \rho_B = \frac{\iden_A}{d_A} \otimes \rho_B, 
\end{equation}
and so $\E_A$ must be unital. We conclude that encodings into resources destroyed by the $G$-twirling channel $\G$ over all unitaries on the $A$ subsystem of the composite $AB$ system correspond to local unital encodings on the $A$ subsystem.

\subsection{Procedure for collective twirling}

\label{app:collective}

In this appendix we discuss how to find a pure state $\ket{\psi^*_{1\dots n}}$
that is mapped to a maximally mixed state by a resource destroying map 
$\mathcal{G}_{\mathrm {col}}$ given by Eq.~\eqref{eq:G_col}. For a given operator $A$, first using the results of Collins and \'S{}niady~\cite{collins2006integration} and then employing the explicit expressions provided by Audenaert~\cite{audenaert2006digest}, we have
\begin{equation}
\mathcal{G}_{\mathrm {col}}(A)=
\frac{1}{n!} \sum_{\pi \in S_n} 
\tr{A P_{\pi}}P_{\pi^{-1}} \sum_{\lambda \vdash n} 
\frac{f^{\lambda}}{s_{\lambda}(1^{ \times  d})} P^{\lambda}.
\end{equation}
In the above $P_{\pi}$ is an operator matrix responsible for permutations of 
subsystems according to a given permutation $\pi$ and $f^\lambda$ is the number of 
standard Young tableaux with shape given by a partition $\lambda$ of $n$ (denoted $\lambda \vdash n$). Next, 
$s_\lambda(1^{ \times  d})$ is a Schur polynomial related to partition 
$\lambda$ evaluated at a point 
\begin{equation}
1^{ \times  d}:=\underbrace{1,1,\dots,1}_d.
\end{equation}
Finally, operators $P^\lambda$ are orthogonal projectors indexed by $\lambda$, which form an orthogonal set and add up to identity operator. These projectors can be defined using permutation matrices $P_\pi$ and the character $\chi^{\lambda}(\pi)$ of permutation $\pi$ (in irreducible representation of symmetric group labelled by partition $\lambda$),
\begin{equation}
P^{\lambda} = \frac{f^{\lambda}}{n!} \sum_{\pi \in S_n} \chi^{\lambda}(\pi) 
P_{\pi}.
\end{equation}
Crucially, observe that projectors $P^\lambda$ are invariant under the action 
of $\mathcal{G}_{\mathrm {col}}$,
\begin{equation}
\mathcal{G}_{\mathrm {col}}(P^\lambda) = P^\lambda.
\end{equation}
For detailed definitions of components and factors above we refer the reader 
to Ref.~\cite{audenaert2006digest}.

In order to construct $\ket{\psi^*_{1\dots n}}$ one can look for states $\ket{x^\lambda}$, which belong to the non-zero eigenspace of $P^\lambda$, i.e.
\begin{equation}
P^\lambda\ket{x^\lambda} = \ket{x^\lambda},
\end{equation}
and such that under the action of $\mathcal{G}_{\mathrm {col}}$ they are mapped onto the full subspace,
\begin{equation}
\mathcal{G}_{\mathrm {col}}(\ketbra{x^\lambda}{x^\lambda}) = 
\frac{P^\lambda}{\tr{P^\lambda}}.
\end{equation}
Then, the following linear combination of such states,
\begin{equation}
\ket{x} := \sum_{\lambda\vdash n}  \sqrt{\frac{\tr{P^\lambda}}{d^n}} \ket{x^\lambda}
\end{equation}
would give us the desired state $\ket{\psi^*_{1\dots n}}$, because
\begin{equation}\label{key}
\mathcal{G}_{\mathrm {col}}(\ketbra{x}{x}) = \sum_{\lambda\vdash n}\frac{P^\lambda }{d^n} = 
\frac{\iden_{d^n}}{d^n} .
\end{equation}

As a particular example consider the case of three qubits, $d=2$ and $n=3$. We then have $P^{\{1,1,1\}} = 0$ and 
\begin{subequations}
\begin{align}\label{key}
\!\!\!  P^{\{2,1\}} &= 
\frac{1}{3}\!\left(
\begin{matrix}
0 & 0 & 0 & 0 & 0 & 0 & 0 & 0 \\
0 & 2 & -1 & 0 & -1 & 0 & 0 & 0 \\
0 & -1 & 2 & 0 & -1 & 0 & 0 & 0 \\
0 & 0 & 0 & 2 & 0 & -1 & -1 & 0 \\
0 & -1 & -1 & 0 & 2 & 0 & 0 & 0 \\
0 & 0 & 0 & -1 & 0 & 2 & -1 & 0 \\
0 & 0 & 0 & -1 & 0 & -1 & 2 & 0 \\
0 & 0 & 0 & 0 & 0 & 0 & 0 & 0 
\end{matrix}
\right)\!,\! \\
\!\!\!  P^{\{3\}} &= 
\frac{1}{3}\!\left(
\begin{matrix}
1 & 0 & 0 & 0 & 0 & 0 & 0 & 0 \\
0 & 1 & 1 & 0 & 1 & 0 & 0 & 0 \\
0 & 1 & 1 & 0 & 1 & 0 & 0 & 0 \\
0 & 0 & 0 & 1 & 0 & 1 & 1 & 0 \\
0 & 1 & 1 & 0 & 1 & 0 & 0 & 0 \\
0 & 0 & 0 & 1 & 0 & 1 & 1 & 0 \\
0 & 0 & 0 & 1 & 0 & 1 & 1 & 0 \\
0 & 0 & 0 & 0 & 0 & 0 & 0 & 1 
\end{matrix}
\right)\!.\!
\end{align}
\end{subequations}
Now, if we take
\begin{subequations}
\begin{align}
\ket{x^{\{2,1\}}} &= \frac{1}{2 \sqrt{3}}  
( 0,-2,1,-\sqrt{3},1,\sqrt{3},0,0)^\top ,\\
\ket{x^{\{3\}}} &=\frac{1}{\sqrt{6}}  (0,1,1,1,1,1,1,0)^\top,
\end{align}
\end{subequations}
and construct
\begin{equation}
\ket{x} = \frac{1}{\sqrt{2}}( \ket{x^{\{2,1\}}}
+ \ket{x^{\{3\}}}),
\end{equation}
we obtain 
\begin{equation}
\mathcal{G}_{\mathrm {col}}(\ketbra{x}{x}) = \frac{\iden_8}{8}.
\end{equation}

\bibliographystyle{IEEEtran}

\begin{IEEEbiographynophoto}{Kamil Korzekwa} is an Assistant Professor and a Junior Group Leader at the Jagiellonian University in Krak\'{o}w, Poland. He received his M.Sc.\ in Nanoengineering from Wroc{\l}aw University of Technology in 2012 and graduated with a Ph.D.\ in Physics from Imperial College London in 2016. Previously, he has been a Research Fellow at the School of Physics at the University of Sydney (2017-2019) and at the International Centre for the Theory of Quantum Technologies at the University of Gda\'{n}sk (2019).

His research lies at the interface of open quantum dynamics, quantum information and resource theories, and is mainly focused on classical-quantum divide arising within various thermodynamic and information processing tasks. This includes exploring protocols that may exhibit quantum advantage, developing various approaches to capture the notions of reversibility and irreversibility in the quantum regime, and studying structural differences between classical and quantum theories.

\end{IEEEbiographynophoto}

\begin{IEEEbiographynophoto}{Zbigniew Pucha\l{}a} is an Associate Professor at the Institute of Theoretical and Applied Informatics, Polish Academy of Sciences and	Assistant Professor at the Faculty of Physics, Astronomy and Applied Computer Science, Jagiellonian University in Kraków.
	
He received an M.Sc.\ in Mathematics from the University of Wroc\l{}aw (Poland) in 2003, graduated with a PhD \ in Mathematics at the Institute of Mathematics at the University of Wroc\l{}aw in 2007 and obtained a habilitation in computer science from the Faculty of	Automatic Control, Electronics and Computer Science at the Silesian University of Technology in Gliwice (Poland) in 2014.
	
His research interests lie in the intersection of mathematics and quantum information theory. The main focus is on the foundations of the theory of quantum information, for example, the study of quantum state and channels discrimination, entropic uncertainty relations and geometrical methods in quantum information and computations.
\end{IEEEbiographynophoto}

\begin{IEEEbiographynophoto}{Marco Tomamichel} (M'13--SM'16) is an Associate Professor at the Department of Electrical and Computer Engineering and a Principal Investigator at the Centre for Quantum Technologies at the National University of Singapore. He received a M.Sc.\ in Electrical Engineering and Information Technology from ETH Zurich (Switzerland) in 2007 and graduated with a Ph.D.\ in Physics at the Institute of Theoretical Physics at ETH Zurich in 2012. Before commencing his current position, he has been a Lecturer at the School of Physics at the University of Sydney and a Senior Lecturer and Associate Professor at the University of Technology Sydney.
	
His research interests lie in the intersection of information theory, computer science and quantum physics. The main focus is on the mathematical foundations of the theory of quantum information, for example the study of entropy and other information measures and their applications to questions in quantum communication, computation and cryptography in the setting where the available resources are limited.
\end{IEEEbiographynophoto}

\begin{IEEEbiographynophoto}{\bf Karol {\.Z}yczkowski} was born in Cracow, Poland in 1960. He received the Ph.D. degree in theoretical physics at the Jagiellonian University in 1987 and obtained there the habilitation in 1994. He was a Humboldt Fellow at the University of Essen (1989/90), a Senior Fulbright Fellow at the University of Maryland (1997/98) and held a one-year research position at the Perimeter Institute in Waterloo (2004/2005).

He is a professor of physics at the Jagiellonian University in Cracow and at the Center for Theoretical Physics, Polish Academy of Science, in Warsaw. His research interests include quantum mechanics, entanglement, quantum information processing, chaos and nonlinear dynamics, random matrices and applied mathematics, mathematical social choice including the theory of elections. Co-author of a book {\sl Geometry of Quantum States}. Recently he works on quantization of combinatorial structures and their applications in quantum information.
	
Prof. {\.Z}yczkowski is a Member of Academia Europaea (2014), Polish Academy of Learning (2021) and Polish Academy of Sciences (2022). Since 2019 he serves as Director of the National Center for Quantum Information in Gda{\'n}sk.
\end{IEEEbiographynophoto}

\end{document}